\documentclass[a4paper,11pt]{article}
\pdfoutput=1
\usepackage{cite}
\usepackage[all]{xy}
\usepackage{amsmath,amsfonts,amssymb,amsthm,epsfig,amscd,comment,latexsym,psfrag}
\usepackage{CJK}
\usepackage{mathrsfs}
\usepackage{nicefrac,xspace,tikz}
\usepackage{arydshln}
\usepackage{stmaryrd}
\usepackage{graphicx,bm}
\usepackage{amscd}
\usetikzlibrary{arrows}

\def\be{\begin{eqnarray}}
\def\ee{\end{eqnarray}}
\def\nn{\nonumber}
\def\p{\partial}
\def\tr{{\rm tr}\,}
\def\Tr{{\rm Tr}\,}

\definecolor{red}{rgb}{1,0,0}
\definecolor{orange}{rgb}{1,0.5,0}
\definecolor{violet}{rgb}{0.7,0,1}

\usepackage{tikz}
\usetikzlibrary{arrows,snakes,backgrounds,calc}
\usepackage{ytableau}
\usepackage[vcentermath]{youngtab}

\newcommand{\ad}{ \operatorname{ad} }
\def\h1{ \hat{h}_1}

\makeatletter

\newcommand{\Rmnum}[1]{\expandafter\@slowromancap\romannumeral #1@}
\makeatother
\usepackage{graphicx,amssymb,amsmath,bm,latexsym}
\allowdisplaybreaks

\begin{document}
\begin{center}

\begin{center}
\hfill FIAN/TD-05/23\\
\hfill IITP/TH-03/23\\
\hfill ITEP/TH-03/23\\
\hfill MIPT/TH-03/23
\end{center}

\bigskip

\bigskip

{\Large\bf  $(q,t)$-deformed (skew) Hurwitz $\tau$-functions}\vskip .2in
{\large Fan Liu$^{a,}$\footnote{liufan-math@cnu.edu.cn},
A. Mironov$^{b,c,d,}$\footnote{mironov@lpi.ru;mironov@itep.ru},
V. Mishnyakov$^{b,c,e,}$\footnote{mishnyakovvv@gmail.com},
A. Morozov$^{c,d,e,}$\footnote{morozov@itep.ru},\\
A. Popolitov$^{c,d,e,}$\footnote{popolit@gmail.com},
Rui Wang$^{f,}$\footnote{wangrui@cumtb.edu.cn},
Wei-Zhong Zhao$^{a,}$\footnote{zhaowz@cnu.edu.cn}} \vskip .4in
$^a$ {\small {\it School of Mathematical Sciences, Capital Normal University,
Beijing 100048, China}} \\
$^b$ {\small {\it Lebedev Physics Institute, Moscow 119991, Russia}}\\
$^c$ {\small {\it NRC ``Kurchatov Institute", 123182, Moscow, Russia}} \\
$^d$ {\small {\it Institute for Information Transmission Problems, Moscow 127994, Russia}}\\
$^e$ {\small {\it MIPT, Dolgoprudny, 141701, Russia}}\\
$^f$ {\small {\it Department of Mathematics, China University of Mining and Technology, Beijing 100083, China}}

\bigskip

\bigskip

\begin{abstract}

We follow the general recipe for constructing commutative families of $W$-operators, which provides Hurwitz-like expansions in symmetric functions (Macdonald polynomials), in order to obtain a difference operator example that gives rise to a $(q,t)$-deformation of the earlier studied models. As before, a key role is played by an appropriate deformation of the cut-and-join rotation operator. We outline its expression both in terms of generators of the quantum toroidal algebra and in terms of the Macdonald difference operators.

\end{abstract}
\end{center}


\section{Introduction}

An interesting class of non-perturbative partition functions is defined
by the action of operators ${\hat W}\{p\}$ from ${W}_{1+\infty}$ algebra
on a trivial vacuum state:
\be\label{PF}
Z= e^{\hat{ W}}\cdot e^{\sum_k \frac{g_kp_k}{k}}
\ee
This $W$-representation was originally found for two matrix models:
for the simplest Gaussian Hermitian matrix model  \cite{MSh} and for the Kontsevich model \cite{Al},
and was later generalized \cite{Alexandrov,Al2,Max,MMMR,MMM} to other models.
In \cite{WLZZ} (see also \cite{WLZZ1}), it was proposed to consider such partition functions {\it per se},
without a reference to matrix models,
and this produced a whole double-parametric series of theories
associated with different operators from  ${W}_{1+\infty}$, \cite{MMCal,MMMP}.
They are naturally divided in two branches, called ``negative" and ``positive"
and associated with generators (of Borel subalgebras) of ${W}_{1+\infty}$ algebra that we denote $\hat E$ and $\hat F$ correspondingly:
\begin{equation}\label{Wop}
	\begin{split}
	\hat  W_{-n}^{(m)} =& {1\over (n-1)!}\ {\rm ad}_{\hat E_{m+1}}^{n-1} \hat E_{m}
	\\
	\hat  W_{n}^{(m)} =& {(-1)^{n-1}\over (n-1)!}\ {\rm ad}_{\hat F_{m+1}}^{n-1} \hat F_{m}
	\end{split}
\end{equation}

The simplest of these are $\hat E_0 = p_1$ and $F_0=\frac{\p}{\p p_1}$.

Formula (\ref{PF}) describes the partition function associated with the positive branch:
\be\label{pos}
Z_+^{(m)}(g,p)= e^{\sum{\bar p_n\hat W_{n}^{(m)}(p)\over n}}\cdot e^{\sum_k \frac{g_kp_k}{k}}
\ee
This partition function can be rewritten as a power series
\be\label{Zp}
Z_+^{(m)}(g,p)=\sum_{\lambda,\mu}\prod_{a=1}^m\left({\prod_{i,j\in \lambda}(N_a+j-i)\over\prod_{i,j\in \mu}(N_a+j-i)}\right)
S_{\lambda/\mu}\{\bar p_k\}S_\lambda\{g_k\}S_\mu\{p_k\}
\ee
where $S_\lambda\{p_k\}$ is the Schur function labeled by partition $\lambda$, which is a graded polynomial of $p_k$, and $S_{\lambda/\mu}\{p_k\}$ is the skew Schur function \cite{Macdonald}. Here $N_a$ are just parameters. The expansion (\ref{Zp}) means that $Z_+^{(m)}(g,p)$ is a $\tau$-function of the KP hierarchy of the skew hypergeometric type \cite{Mironov:2023mve}.

The partition function associated with the negative branch is obtained from the positive one upon putting all $p_k=0$:
\be
Z_-^{(m)}(p)=Z_+^{(m)}(g,p)\Big|_{p_k=0}
\ee
It can be still presented in the form (\ref{PF}):
\be\label{neg}
Z_-^{(m)}(g)= e^{\sum{\bar p_n \hat  W_{-n}^{(m)}(g)\over n}}\cdot 1
\ee
In this case, it is possible to act with the $W$-operator on the trivial state and still to get a non-trivial answer,
while for the positive branch some $g_k$ should be kept non-vanishing in order
to get $Z_+^{(m)}\neq 1$. The partition function (\ref{neg}) can be written as
\be\label{Zm}
Z_-^{(m)}(g)=\sum_\lambda\prod_{a=1}^m\prod_{i,j\in \lambda}(N_a+j-i)S_{\lambda}\{\bar p_k\}S_\lambda\{g_k\}
\ee
which means it is a $\tau$-function of the KP hierarchy of the hypergeometric type \cite{GKM2,OS,O,AMMN}.

Such partition functions possess many interesting properties,
including integrability \cite{UFN1,UFN2,UFN3,UFN4,UFN5} and superintegrability \cite{MMsi}.
There are also good chances for the AMM/EO topological recursion \cite{toprec1,toprec2,toprec3,toprec4,toprec5,toprec6},
at least one can define the plausible spectral curves \cite{MMsc},
amusingly, for the positive branch, this is at expense of the standard relation
to topology through the $1/N$ expansion.
Moreover, they typically possess matrix model representations like
\cite{Alexandrov,Alexandrov:2022whk,Mironov:2023pnd,Mironov:2023mve}
\be
Z_-^{(1)}(g)=\int dXdY  e^{-\Tr XY + \sum_{k} \frac{g_k}{k}\Tr X^k + \sum_k \frac{\bar p_k}{k} \Tr Y^k},
\ee
or
\be
Z_+^{(1)}(g,p)=\int dXdY  e^{-\Tr XY + \Tr \Lambda Y  + \sum_{k} \frac{g_k}{k}\Tr X^k + \sum_k \frac{\bar p_k}{k} \Tr Y^k},
\ \ \ \ \ \ p_k = \tr \Lambda^k
\ee
for the negative and positive branches respectively.
These directions are, however, rather tricky and can need new insights.
\\\\

What is straightforward,
one can perform a  $(q,t)$-deformation of these models,
which substitutes the Schur functions in superintegrability expansions by the Macdonald polynomials \cite{Macdonald}
as suggested in \cite{Morozov:2018eiq,Morozov,Cassia} (see also \cite{Cassia1}).
This is somewhat technical, still such a deformation is important for further application
to $5d$ and $6d$ string theories.
Not less important is that such generalizations are associated with lifting of $W_{1+\infty}$
to generic quantum toroidal algebras, and such study sheds new light on the still-not-very-well-known formalism,
which is going to play a crucial role in the near future.
This paper is the first attempt of such presentation.
It is not yet pedagogical enough, but a long way begins from the first step.
We believe that the models in question are indeed the archetypical and the most fundamental ones,
and this research will have many continuations and improvements in various directions (see first steps in \cite{WLY}).

\paragraph{Notation.} Throughout the text, we work with the Macdonald polynomials $P_\lambda$ as with symmetric polynomials of variables $x_i$, or deal with them as with graded polynomials of the power sums $p_k=\sum_i x_i^k$. We denote them as $P_\lambda(x;q,t)$ and $P_\lambda\{p_k\}$, correspondingly.

We use the notation $\langle \cdot ,\cdot \rangle_{(q,t)}$ for the Macdonald scalar product \cite[p.309, Eq.(2.2)]{Macdonald}.

We also define the quantity
\footnote{Note that the $(q,t)$ deformation of the specialization locus we use here slightly differs from the one used in \cite{Morozov:2018eiq,Cassia} (note that, in the first of these papers, the notation is $(q,t)\to(q^2,t^2)$), however the Macdonald polynomials evaluated at these points differ only in normalization:
\begin{equation}
	\frac{P_\lambda\left(p_k=n \frac{\left(q^{-1 / 2}-q^{1 / 2}\right)^{k / n}}{t^{-k / 2}-t^{k / 2}} \delta_{k \mid n}\right)}{P_\lambda\left(p_k=n \frac{\left(1-q^n\right)^{k / n}}{1-t^k} \delta_{k \mid n}\right)}= \left(t q^{1 / n}\right)^{|\lambda| / 2}\left(\frac{1-q}{1-q^n}\right)^{|\lambda| / n}
\end{equation}
when $|\lambda|$ is divisible by $n$ (otherwise, the Macdonald polynomials are zero at all these loci).
  }
\be
\delta^{*}_{k,n}=\dfrac{(1-q^n)^{k/n}}{1-t^k}\ \sum_{j=0}^{n-1}e^{2\pi i kj\over n}=
\left\{\begin{array}{cl}
n\dfrac{(1-q^n)^{k/n}}{1-t^k}&\hbox{if }k=0\hbox{ mod }n\cr
0&\hbox{otherwise}
\end{array}\right.
\ee

\bigskip

\section{$(q,t)$-deformation of partition functions and $W$-representation}

\subsection{Commuting families in $W_{1+\infty}$ and the $\hat{O}$ automorphism}

Let us briefly recollect the main points of the constructions presented in \cite{Mironov:2023mve,Mironov:2023pnd}. The key idea is that, in order to generate partition functions \eqref{PF}, one needs a commutative family of $W$-operators with a certain action on characters. To construct such a family, one follows the steps:
\begin{enumerate}
	\item Consider the commutative family of operators $p_k$ or $\frac{\partial}{\partial p_k}$.
	\item Introduce the operator $\hat{O}(N)$, which we called the cut-and-join rotation operator, with the following action on characters:
	\begin{equation}
		\hat{O}(N) S_R =\left( \prod_{(i,j) \in R}(N+j-i) \right) S_R
	\end{equation}
	\item Construct by conjugation from $p_k$ or from $\dfrac{\partial}{\partial p_k}$ operators that are commuting and have the correct action on characters:
	\begin{equation}\label{Wop}
	\begin{split}
				W^{(m)}_{-n}(\vec{N}) &= \hat{O}(\vec{N}) \cdot p_n \cdot \hat{O}(\vec{N})^{-1}
				\\
				W^{(m)}_n(\vec{N}) &= \hat{O}(\vec{N})^{-1} \cdot n\dfrac{\partial}{\partial p_n} \cdot \hat{O}(\vec{N})
	\end{split}
	\end{equation}
where
\begin{equation}
	\hat{O}(\vec{N})=\prod_{l=1}^{m} \hat{O}(N_l)
\end{equation}
	\item The cut-and-join rotation operator $\hat O(N)$ is represented in term of the well-known ordinary cut-and-join operators in an explicit, albeit rather complicated way \cite{Alexandrov}. Therefore, if one wishes an explicit expression for operators \eqref{Wop} in terms of $p_k$-variables, another realization is available in terms of iterated commutators in the $W_{1+\infty}$ algebra:
	\begin{equation}
		\begin{split}
			\hat  W_{-n}^{(m)} =& {1\over (n-1)!}\ {\rm ad}_{\hat E_{m+1}}^{n-1} \hat E_{m}
			\\
			\hat  W_{n}^{(m)} =& {(-1)^{n-1}\over (n-1)!}\ {\rm ad}_{\hat F_{m+1}}^{n-1} \hat F_{m}
		\end{split}
	\end{equation}
with $E_{m}=\ad^{m}_{W_0} p_1, F_{m}=\ad^m_{W_0} \left( \frac{\partial}{\partial p_1} \right)$ and $W_0$ is the simplest cut-and-join operator, sometimes also labelled as $W_{[2]}$
\cite{Mironov:2009cj,Mironov:2010yg}.
\end{enumerate}
Having made these steps, the explicit character realization of the (skew) $\tau$-functions is completely straightforward as outlined in \cite{Mironov:2023mve} (see specifically Appendix C in \cite{Mironov:2023pnd}).

\subsection{$(q,t)$-deformed partition functions}

Now we construct $(q,t)$-deformed partition functions along the line of the previous subsection. It is well known that, in the case of difference operators, the correct substitutions for the Schur functions are the Macdonald polynomials. In this case, a proper generalization of the content of the partition box is given by
\begin{equation}
	(N+j-i) \rightarrow {1-q^{j-1}t^{-i+1+N}\over 1-q}
\end{equation}
The partition functions that we would like to construct are built using the Cauchy identity for the Macdonald polynomials:
\begin{equation}
	\exp\left(\sum \dfrac{1-t^n}{1-q^n} \dfrac{p_k \bar{p_k}}{k} \right) = \sum_\lambda P_\lambda\{p\} Q_\lambda\{\bar{p}\}
\end{equation}
where $Q_\lambda$ are the dual Macdonald polynomials:
\begin{equation}
	Q_\lambda = \dfrac{P_\lambda}{ \quad \Big\langle P_\lambda , P_{\lambda} \Big\rangle_{(q,t)}}
\end{equation}
and the Macdonald norm square is explicitly
\begin{equation}
	\Big\langle P_\lambda , P_{\mu} \Big\rangle_{(q,t)} = \delta_{\lambda,\mu} \prod_{(i,j) \in \lambda} \dfrac{t^{-\lambda_j^T+i}q^{-\lambda_i+j-1}-t^{\lambda_j^T-i}q^{\lambda_i-j+1}}{t^{-\lambda_j^T+i-1}q^{-\lambda_i+j}-t^{\lambda_j^T-i+1}q^{\lambda_i-j}}
\end{equation}
Therefore, taking into account the deformation of the content functions, we are looking for the following partition function generalizing (\ref{Zp}):
\begin{equation}\label{zpM}
\boxed{
	Z_{+}^{(q,t)}(\vec{N};\bar{p},p,g)=\sum_{\lambda,\mu} \left(\dfrac{1}{(1-q)^{m(|\lambda|-|\mu|)}} \cdot \dfrac{\prod\limits_{l=1}^{m} \prod\limits_{(i,j) \in \lambda} (1-q^{j-1}t^{-i+1+N_l}) }{  \prod\limits_{l=1}^{m} \prod\limits_{(i,j) \in \mu} (1-q^{j-1}t^{-i+1+N_l})} \right)  Q_\lambda\{g_k\} P_\mu\{p_k\} P_{\lambda/\mu}\{\bar{p}_k\}
}
\end{equation}
At this point, let us note that the only example of matrix/eigenvalue model realizations of such partition functions known so far is the $(q,t)$-deformed Gaussian matrix model \cite{Morozov:2018eiq,Cassia}, which corresponds to the very degenerate case of $p_k=0, \,  \bar{p}_k=\delta^{*}_{k,2}$.
In this case, the partition function is
\begin{equation}
	Z^{(q,t)}_{\text{Gaussian}}(q,t,N) = \sum_\lambda \prod\limits_{(i,j) \in \lambda} \dfrac{1-q^{j-1}t^{-i+1+N}}{1-q}  P_\lambda \{\delta^{*}_{k,2} \} Q_\lambda \{g\}
\end{equation}

\subsection{Commuting families of difference operators}

As in the non-deformed case, in order to generate the partition functions, we introduce the operator
\begin{equation}
	\hat{\mathbf{O}}(q,t|N) P_\lambda = \left( \prod_{(i,j) \in \lambda} \dfrac{1-q^{j-1}t^{-i+1+N}}{1-q} \right)P_\lambda
\end{equation}
Hereafter, we represent the $(q,t)$-deformed operators in ``bold" to distinguish them from their non-deformed counterparts.

Then, the desired $W$-operators can be constructed as follows:
\begin{equation}\label{WqtO}
	\begin{split}
				\mathbf{W}^{(m)}_{-n}(q,t|\vec{N}) &= \hat{\mathbf{O}}_m(q,t|\vec{N}) \cdot p_n\cdot  \left(\hat{\mathbf{O}}_m(q,t|\vec{N}) \right)^{-1}
		\\
		\mathbf{W}^{(m)}_n(q,t|\vec{N}) &=  \hat{\mathbf{O}}_m(q,t|\vec{N})^{-1}\cdot n{1-q^n\over 1-t^n}\dfrac{\partial}{\partial p_n}  \cdot\left(\hat{\mathbf{O}_m}(q,t|\vec{N}) \right)
	\end{split}
\end{equation}
where:
\begin{equation}
	\hat{\mathbf{O}_m}(q,t|\vec{N})=\prod_{l=1}^{m} \hat{\mathbf{O}}(q,t|N_l)
\end{equation}
\\\\
Clearly these families of operators are commuting, which is easily seen using their definition \eqref{WqtO}:
\begin{equation}
	\begin{split}
			&[	\mathbf{W}^{(m)}_{-n_1}(q,t|\vec{N})],	\mathbf{W}^{(m)}_{-n_2}(q,t|\vec{N}) ] =0
			\\
			&[	\mathbf{W}^{(m)}_{n_1}(q,t|\vec{N})],	\mathbf{W}^{(m)}_{n_2}(q,t|\vec{N}) ] =0
	\end{split}
\end{equation}
Therefore, given such operators, the $W$-representations for the $(q,t)$-deformed partition functions are:
\begin{equation}\label{zmu}
	\boxed{Z_{-}^{(q,t)}(\vec{N};g,p) = \exp \left\{ {\sum_{n \geq 1}\frac{1-t^{n}}{1-q^{n}} \frac{	\mathbf{W}^{(m)}_{-n}(q,t|\vec{N})}{n} g_{n}}\right\} \cdot 1} \nonumber
\end{equation}
which is equal to
\begin{equation}
\boxed{
Z_{-}^{(q,t)}(\vec{N};g,p) =\sum_{\lambda}\frac{1}{(1-q)^{m|\lambda|}} \prod\limits_{l=1}^{m} \prod\limits_{(i,j) \in \lambda} (1-q^{j-1}t^{-i+1+N_l})
	P_{\lambda}\left\{g\right\} Q_{\lambda}\left\lbrace p\right\rbrace=
}
 \end{equation}
$$
=\sum_{\lambda} \prod_{l=1}^{m}\frac{P_{\lambda} \left\{\frac{1-t^{k N_l}}{1-t^k}\right\}}
	{P_{\lambda}\left\{\delta_{k,1}^{*}\right\}}
	P_{\lambda}\left\lbrace g\right\rbrace Q_{\lambda}\left\lbrace p\right\rbrace.
$$
for the negative branch, and
\begin{equation}
\boxed{
	Z^{(q,t)}_{+}(\vec{N};\bar{p},p,g) =  \exp \left\{ {\sum_{n \geq 1}\frac{1-t^{n}}{1-q^{n}} \frac{	\mathbf{W}^{(m)}_{-n}(q,t|\vec{N})}{n} \bar{p}_{n}}\right\} \cdot \exp\left(\sum \dfrac{1-t^n}{1-q^n} \dfrac{p_k g_k}{k} \right)
=(\ref{zpM})
}
\end{equation}
for the positive branch. These formulas are the $(q,t)$-deformations of formulas (\ref{neg}), (\ref{Zm}), (\ref{pos}), (\ref{Zp}).

\subsection{Explicitly constructing $\mathbf{W}^{(m)}_{\pm n}$}

In the non-deformed case, the realization of the $\hat O(N)$ operator in terms of $p_k$-variables is complicated, and even more so in the $(q,t)$-deformed case. In the next section, we express it as an infinite product of some more familiar operators, the higher Ruijsenaars Hamiltonians \cite{Rui1,Rui2} (or the Macdonald difference operators \cite[p.315, Eq.(3.4)]{Macdonald}). Hence, here we develop another approach: we introduce a scheme which allows one to construct the operators $\mathbf{W}^{(m)}_{\pm n}$ iteratively in the very explicit form as difference operators. It is a $(q,t)$-deformation of the scheme explained in \cite{Mironov:2023pnd}, however, it is not immediate and requires some modification.
\\\\
A key ingredient of iterative formulas for ${W}^{(m)}_{\pm n}$ proposed in \cite{Mironov:2023pnd} is the diagonal cut-and-join operator $W_0$. Luckily, it has a straightforward $(q,t)$-deformation, which is now realized as the Macdonald difference operator in the $x$ variables.
The Macdonald difference operator $D_{N}^{(1)}$ is given by \cite{Macdonald}
\begin{eqnarray}
	D_{N}^{(1)}&=&\sum_{i=1}^{N}(\Delta_N^{-1}T_{t,i}\cdot\Delta_N)T_{q,i} \nonumber \\ &=&\sum_{i=1}^{N} \left( {\prod_{j=1,j\neq i}^{N} \frac{tx_{i}-x_{j}}{x_{i}-x_{j}}}\right)T_{q,i},
\end{eqnarray}
where $\Delta_N=\prod_{i<j}^N(x_i-x_j)$ is the Vandermonde determinant, and $T_{\xi,i}=\xi^{x_i{\p \over \p x_i}}$ is the operator of dilatation of the $i$-th variable $x_i$: $x_i\to \xi x_i$.
It acts on the Macdonald polynomial $P_{\lambda}(x;q,t)$ as
\begin{equation}\label{dn1}
	D_{N}^{(1)}P_{\lambda}(x;q,t)=\left(\sum_{i= 1}^{N}q^{\lambda_{i}}t^{N-i}\right)P_{\lambda}(x;q,t).
\end{equation}
In order to obtain the desired functions of $q^j t^{-i}$, we define a rescaled and shifted operator, which is the $(q,t)$-deformed cut-and-join operator
\begin{equation}\label{w0}
	W_0\longrightarrow\mathcal{W}_0(q,t|N)=\frac{1}{1-q}\hat{L}_0+\frac{t^{N+1}}{(1-q)^2}\left(t^{-N} D_{N}^{(1)}+\frac{1-t^{-N}}{1-t} \right),
\end{equation}
where $\hat{L}_0=\sum\limits_{i=1}^{N} x_i\dfrac{\partial}{\partial x_i}$ is the Euler operator.
Then, one has
\begin{eqnarray}\label{w0p}
		\mathcal{W}_0(q,t|N)P_{\lambda}(x;q,t)=\sum_{(i,j)\in \lambda}C^{(q,t)}(N;i,j)P_{\lambda}(x;q,t),
\end{eqnarray}
where $C^{(q,t)}(N;i,j):=\frac{1-q^{j-1}t^{N+1-i}}{1-q}$.
\\\\
Just as in the non-deformed case, we further define the operator
\begin{equation}
\mathbf{E}_1(q,t|N)=[\mathcal{W}_0(q,t|N),p_{1}]
\end{equation}
which has the following action on the Macdonald polynomials
 $$ \mathbf{E}_1(q,t|N)P_{\lambda}=\sum_{\lambda+\Box}C^{(q,t)}(N;i_{\Box},j_{\Box})
    \Big\langle p_{1}P_{\lambda},Q_{\lambda+\Box} \Big\rangle _{(q,t)} P_{\lambda+\Box}$$
The recursive commutator formulas require some adjustments.  A naive generalization of formulas \eqref{Wop} does not work. For example, a simple check shows that
\begin{equation}
	[[\mathcal{W}_0,\mathbf{E}_1],\mathbf{E}_1] \neq \mathbf{W}_{-2}^{(1)}
\end{equation}
Instead, we construct the operators $\mathbf{W}_{-n}^{(1)}(q,t|N)$ by the following commutators:
\begin{equation}\label{W-nn}
  \frac{(1-q)^{n}}{1-q^{n}}[\mathbf{W}_{-n}^{(1)}(q,t|N) ,D_{N}^{(1)}]= \overset{n}{\overbrace{[\mathbf{E}_1(q,t|N),[\dots,[\mathbf{E}_1(q,t|N)}},D_{N}^{(1)}]\dots]],
\end{equation}
which is a counterpart of a similar recursive relation involving $p_k$-variables:
\begin{equation}\label{recu}
  \frac{(1-q)^{n}}{1-q^{n}}[p_{n},D_{N}^{(1)}]=
  \overset{n}{\overbrace{[p_{1},[\dots,[p_{1}}},D_{N}^{(1)}]\dots]].
\end{equation}
As an incidental note let us mention that, for generic $p_s$, this recursive relation is
\begin{equation}\label{recus}
  \frac{(1-q^s)^{n}}{1-q^{ns}}[p_{ns},D_{N}^{(1)}]=
  \overset{n}{\overbrace{[p_{s},[\dots,[p_{s}}},D_{N}^{(1)}]\dots]].
\end{equation}
Both relations follow from the simple commutation relation:
\begin{equation}
	\left[x_i^m T_{q,i} , \sum_{k=1}^N x_k^n \right] = x_i^{m+n} (q^n-1) T_{q,i}
\end{equation}
Thus, one has
\begin{eqnarray}
  \sum_{n\ge 1}{1\over n}\overset{n}{\overbrace{[p_{s},[\dots,[p_{s}}},D_{N}^{(1)}]\dots]]&=&\sum_{k\ge 1} {1\over k}\frac{1-t^k}{1-q^k}\frac{s(1-q^s)^{k/s}} {1-t^k}\delta_{k|s}[p_{k},D_{N}^{(1)}] \nonumber\\
  &=&\sum_{k\ge 1} {1\over k}\frac{1-t^k}{1-q^k}\delta_{k,s}^* [p_{k},D_{N}^{(1)}].
\end{eqnarray}
It appears that the locus $p_k=\delta_{k,s}^*$ is the correct $(q,t)$-deformation of the specializations of  $p_k$-variables, which appears in specific models such as $(q,t)$ matrix models that realize partition functions under consideration \cite{Morozov:2018eiq,Cassia}.
The recursive relations \eqref{W-nn} and \eqref{recu} look differing from their non-deformed counterparts. Moreover, it is not even obvious that they have a well-defined limit at $q=t=1$. We deal with this issue in the next subsection, and, right here, we note that the $W$-operators constructed in such a way have the desired action on characters:
\begin{equation}\label{ngr}
	\mathbf{W}_{-n}^{(1)}(q,t|N)P_{\lambda}=\sum_{\mu = \lambda+\Box_{1}+\cdots+\Box_{n}} \left(
	\prod_{(i,j)\in \mu /\lambda}\frac{1-q^{j-1}t^{N+1-i}}{1-q} \right) \Big\langle p_{n}P_{\lambda},Q_{\mu} \Big\rangle _{q,t} P_{\mu}.
\end{equation}
Indeed, notice that, by construction, one has $\mathbf{W}_{-1}^{(1)}(q,t|N)=\mathbf{E}_1(q,t|N)P_{\lambda}=\mathbf{O} p_{1}\hat{\mathbf{O}}^{-1}$, $[\hat{\mathbf{O}},D_{N}^{(1)}]=0$, and, hence,
\begin{eqnarray}\label{opno}
  \frac{(1-q)^{n}}{1-q^{n}}[\mathbf{W}_{-n}^{(1)}(q,t|N),D_{N}^{(1)}]=\hat{\mathbf{O}} \overset{n}{\overbrace{[p_{1},[\dots,[p_{1}}},D_{N}^{(1)}]\dots]] \hat{\mathbf{O}}^{-1} = \frac{(1-q)^{n}}{1-q^{n}}[\hat{\mathbf{O}} p_{n} \hat{\mathbf{O}}^{-1},D_{N}^{(1)}]
\end{eqnarray}
Clearly, \eqref{W-nn} is defined up to operators commuting with $D^{(1)}_N$. For all such operators, the Macdonald polynomials are the eigenfunctions. Hence, generally $\mathbf{W}_{-n}^{(1)}P_\lambda\to P_\lambda+\oplus_{\mu = \lambda+\Box_{1}+\cdots+\Box_{n}} P_\mu$. However, we need the operator (\ref{ngr}) that has the fixed degree $n$ (i.e. that adds $n$ boxes to the partition). Hence, we choose the diagonal piece to vanish so that $\mathbf{W}_{-n}^{(1)}P_\lambda\to \oplus_{\mu = \lambda+\Box_{1}+\cdots+\Box_{n}} P_\mu$.
\\\\
In completely the same way, we construct the higher operator $\mathbf{W}^{(m)}_{-n}(q,t|\vec{N})$ by
\begin{equation}
	\frac{(1-q)^{n}}{1-q^{n}}[	\mathbf{W}^{(m)}_{-n}(q,t|\vec{N}),D_{N}^{(1)}]=
	\overset{n}{ \overbrace{[\mathbf{E}_m(q,t|\vec{N}),[\dots,[\mathbf{E}_m(q,t|\vec{N})
	} },D_{N}^{(1)} ]\dots]],
\end{equation}
and the operators $\mathbf{E}_m(q,t|\vec{N})$
are given by
\begin{equation}
\mathbf{E}_m(q,t|\vec{N})= \overset{m}{\overbrace{[	\mathcal{W}_0(q,t|N_m),[\dots,[	\mathcal{W}_0(q,t|N_1)}},p_{1}]\dots]].
\end{equation}
\\\\
The negative branch is treated in a similar way. Denote $\hat{A}^*$ the adjoint of $\hat{A}$ with respect to the scalar product $\langle\cdot,\cdot\rangle_{q,t}$, then $(D_N^{(1)})^*=(D_N^{(1)})$ and $(p_n)^*=n\frac{1-q^n}{1-t^n}{\p\over\p p_n}$. We can construct the operators $W_{n}$ as the adjoint of (\ref{W-nn})
\begin{equation}
   \frac{(1-q)^{n}}{1-q^{n}}[D_{N}^{(1)},\mathbf{W}_n(q,t|N)]=[[\cdots[D_{N}^{(1)},\overset{n}{\overbrace{\mathbf{F}_{1}(q,t|N)],\cdots ],\mathbf{F}_{1}(q,t|N)}}],
\end{equation}
 where $\mathbf{F}_{(1)}(q,t|N)=\frac{1-q}{1-t}[\frac{\partial}{\partial p_{1}}, \mathbf{W}_{0}(q,t|N)]$. The higher operators are given by
 \begin{equation}\label{Wpos}
 	\frac{(1-q)^{n}}{1-q^{n}}[D_{N}^{(1)},\mathbf{W}^{(m)}_n(q,t|\vec{N})]=
 	[\cdots[D_{N}^{(1)},\overset{n}{\overbrace{ \mathbf{F}_{m}(q,t|N)],\cdots ], \mathbf{F}_{m}(q,t|N)}}],
 \end{equation}
 where
 \begin{equation}
 \mathbf{F}_{m}(q,t|N)= \frac{1-q}{1-t}[[\cdots [ \frac{\partial}{\partial p_{1}},\overset{m}{\overbrace{ \mathcal{W}_0(q,t|N_1)],\dots],\mathcal{W}_0(q,t|N_m) }}]
 \end{equation}

\subsection{Non-deformed limit}

The new recursive formulas for the $\mathbf{W}^{(m)}_{\pm n}$ operators \eqref{W-nn} and \eqref{Wpos}, which emerge in the $(q,t)$-deformed case look different as compared to the non-deformed formulas of \cite{Mironov:2023pnd}. Hence, taking limit to the non-deformed case requires some care. Here we describe the limiting procedure explicitly.
\\\\
Let us start with \eqref{recu}:
\begin{equation}
	\overset{n}{\overbrace{[p_{1},[\dots,[p_{1}}},D_{N}^{(1)}]\dots]]=	\frac{(1-q)^{n}}{1-q^{n}}[p_{n},D_{N}^{(1)}]
\end{equation}
Coming to the non-deformed case is done by expanding the formulas in $\hbar$,
\begin{equation}
	q=t=e^\hbar
\end{equation}
and then taking the limit of $\hbar=0$. Notice that the r.h.s. of \eqref{recu} is immediately of order $\hbar^{n-1}$. To see that the same is true for the l.h.s. requires some work.
\\\\
First, for technical purposes, define
\begin{equation}
	\hat{h}_{1} = \dfrac{t}{1-q} \left( t^{-N}D_N^{(1)}+\dfrac{1-t^{-N}}{q-t} \right)
\end{equation}
This $\hat{h}_1$ is just another rescaled/shifted version of the first Macdonald Hamiltonian. Its eigenvalues are
\begin{equation}
	\hat{h}_1 P_\lambda = \left( \sum_{(i,j) \in \lambda} q^{j-1} t^{1-i} \right) P_\lambda
\end{equation}
Then, one has
\begin{equation}\label{pqt}
	\overset{n}{\overbrace{[p_{1},[\dots,[p_{1}}},\hat{h}_1]\dots]]=
	\frac{(1-q)^{n}}{1-q^{n}}[p_{n},\hat{h}_1]
\end{equation}
The $\hbar$-expansion of the diagonal operator is given by
\begin{equation}
		        \hat{h}_1= \sum_{n=0}^{\infty} \dfrac{\hbar^{m}}{m!}  V_{(m+2,0)}
\end{equation}
where $V_{(m+2,0)}$ are the degree $0$ and maximal spin $(m+2)$ commuting operators in $W_{1+\infty}$. The simplest representatives are given by $V_{(2,0)}=L_0$ and $V_{(3,0)}=W_0(N=0)$.\\\\
Then, one has in the first non-vanishing order in $\hbar$:
\begin{equation}
	\begin{split}
		\operatorname{ad}_{p_1}^n \hat{h}_1 &= \dfrac{(1-q)^n}{1-q^n} [p_n ,\h1]
		\\ &\Downarrow
		\\
		\sum_{m=0}^{\infty} \dfrac{\hbar^{m}}{m!}  \ad_{p_1}^n V_{(m+2,0)} &=  \dfrac{(-1)^{n+1} \hbar^{n-1}}{n} [p_n,L_0] + o\left(\hbar^{n}\right)
		\\ &\Downarrow
		\\
		\dfrac{\hbar^{n-1}}{(n-1)!} \ad_{p_1}^n V_{(n+1,0)} + o\left(\hbar^{n}\right)&= \dfrac{(-1)^{n+1} \hbar^{n-1}}{n} [p_n,L_0] + o\left(\hbar^{n}\right)
	\end{split}
\end{equation}
In the last line, we used the property
\begin{equation}
	\ad_{p_1}^n V_{(m,0)} =0 \ , \ n \geq m
\end{equation}
which follows from the general commutation relations in the $W_{1+\infty}$-algebra \cite{Bakas:1989xu,Pope1,Pope2,Pope3,Pope4,FKN2,Awata:1994tf} with $p_1=V_{(1,1)}$.
We conclude that formula \eqref{recu} has an appropriate limit to the non-deformed case given by
\begin{equation}
	p_n=\dfrac{(-1)^n}{(n-1)!} \ad^n_{p_1} V_{(n+1,0)}
\end{equation}
The same logic can be applied to \eqref{W-nn}. The expansion at the l.h.s. is organised in such a way that the first $(n-1)$ terms vanish due to the commutation relations in the $W_{1+\infty}$ algebra, while the r.h.s. is proportional to $W_{-n}$ in the leading order.
\section{Constructions of the operator $\hat{\mathbf{O}}(q,t|N)$ }

In order to complete our description of the $(q,t)$-deformation, we now discuss an explicit construction of the operator $\hat{\mathbf{O}}$. It can be done in two different ways: in terms of a Fock representation of the quantum toroidal algebra, and, more explicitly, in terms of the Macdonald difference operators.

\subsection{Operator $\hat{\mathbf{O}}$ and DIM algebra generators}

We first describe a realization of the operators $\hat{\mathbf{O}}$ (and, hence,  $\mathbf{W}_{-n}^{(m)}(q,t|\vec{N})$) in terms of generators of the quantum toroidal algebra.

Let $q_1$, $q_2$ and $q_3$ be formal parameters satisfying $q_1q_2q_3=1$.
The (quantum toroidal) Ding-Iohara-Miki (DIM) algebra $U_{q_1,q_2,q_3}(\hat{\hat{\mathfrak{gl}_1}})$ \cite{DI,Miki} is multiplicatively generated by the central elements $c_1$, $c_2$ and by the elements $e_{\vec{\gamma}}$, with $\vec{\gamma} \in \mathbb{Z}^2\setminus \{(0,0)\}$,
satisfying a set of commutation relations \cite{Feigin,Burban,Zenkevich}. Let
$\mathcal{F}_{q_1,q_2}^{(1,0)}=\mathbb{C}[p_1,p_2,\cdots]$ be the vector space of (graded) polynomials in the variables $p_k$. Denote
$(q_1,q_2,q_3)=(q, t^{-1},q^{-1}t)$ and use the Macdonald polynomials as the basis of polynomials in this vector space. Then, there is a map
\begin{equation}
	f:U_{q_1,q_2,q_3}(\hat{\hat{\mathfrak{gl}}}_1)\rightarrow \mathrm{Aut}(\mathcal{F}_{q,t^{-1}}^{(1,0)})
\end{equation}
defined by manifest action of the generating elements of the algebra $U_{q_1,q_2,q_3}(\hat{\hat{\mathfrak{gl}_1}})$
on the Macdonald polynomials \cite[Eqs.(37)-(45)]{Zenkevich} that gives a representation of $U_{q_1,q_2,q_3}(\hat{\hat{\mathfrak{gl}}}_1)$ on $\mathcal{F}_{q,t^{-1}}^{(1,0)}$, where the superscript $(1,0)$ refers to the values of the two central charges of this algebra. In particular, the action of the Cartan-like elements is

\begin{eqnarray}
	f(e_{(\pm n,0)})P_{\lambda}\left\{p\right\}&=& \pm\left(-{1\over(1-q^{\pm n}) (1-t^{\mp n})}+\sum_{(i,j)\in \lambda}(q^{j-1}t^{1-i})^{\pm n}\right) P_{\lambda}\left\{p\right\}, \quad n>0, \label{en0}
\end{eqnarray}
while
\begin{eqnarray}
	f(e_{(0,n)})&=&n{1\over 1-t^{-n}}{\partial\over \partial p_n},\quad n>0, \nonumber\\
	f(e_{(0,-n)})&=&-(qt^{-1})^{n \over 2}{1\over 1-q^n} p_n,\quad n>0
\end{eqnarray}
Let us define the generating function $\hat{T}_{(1,0)}(u)=
\exp\{-\sum_{n\ge 1}\frac{e_{(n,0)}u^n}{n}\}$, then, from (\ref{en0}), one obtains
\begin{equation}
	f(\hat{T}_{(1,0)}(u))P_{\lambda}\{p\}=\exp\left\{\sum_{n\ge 1}
	\frac{u^n}{n(1-q^n)(1-t^{-n})}\right\}\prod_{(i,j)\in \lambda} (1-uq^{j-1}t^{1-i})P_{\lambda}\{p\}.
\end{equation}
Therefore, the operator $\hat{\mathbf{O}}$ can be expressed as
\begin{equation}
\boxed{
	\hat{\mathbf{O}}=(1-q)^{\hat{E}}\exp\left\{-\sum_{n\ge 1}
	\frac{t^{nN}}{n(1-q^n)(1-t^{-n})}\right\}f(\hat{T}_{(1,0)}(t^N))
}
\end{equation}
where $\hat{E}=\sum_{n\ge1 }np_n{\partial\over\partial p_n}$.
\\\\

Now, using (\ref{WqtO}), one can realize the operators $D^{(1)}_N$, $\mathcal{W}_0(q,t|N)$, $\mathbf{W}_{-n}^{(m)}(q,t|\vec{N})$
and $\mathbf{W}_{n}^{(m)}(q,t|\vec{N})$ in terms of the generators of the DIM algebra in this representation as follows:
\begin{eqnarray}\label{qtW}
	D^{(1)}_N&=&-(1-q)t^{N-1}f(e_{(1,0)})-{t^{-1} \over 1-t^{-1}}, \nonumber\\
	\mathcal{W}_0(q,t|N)&=&{1\over 1-q}(\hat{E}-t^Nf(e_{(1,0)}))- {t^N\over(1-q)^2(1-t^{-1})}, \nonumber\\
	\mathbf{W}_{-n}^{(m)}(q,t|\vec{N})&=&-(qt^{-1})^{n\over 2}(1-q)^n (1-q^n) f\left(\prod_{i=1}^{m}\hat{T}_{(1,0)}(t^{N_i})\cdot e_{(0,-n)}\cdot \prod_{j=1}^{m}\hat{T}_{(1,0)}^{-1}(t^{N_j})\right), \nonumber\\
	\mathbf{W}_{n}^{(m)}(q,t|\vec{N})&=&-t^{-n}\frac{1-q^n}{(1-q)^n} f\left(\prod_{i=1}^{m}\hat{T}_{(1,0)}(t^{N_i})\cdot e_{(0,n)}\cdot \prod_{j=1}^{m}\hat{T}_{(1,0)}^{-1}(t^{N_j})\right).
\end{eqnarray}
Note that one can naturally provide the operators that we constructed with a double grading $\mathbf{d}$ such that
\be
\mathbf{d}\Big(p_n\Big)&=&(0,-n)\ \ \ \ \ \ \ \ \ \ \ \ 
\mathbf{d}\Big({\p\over\p p_n}\Big)=(0,n)\nn\\
\mathbf{d}\Big(\mathcal{W}_0\Big)&=&(1,0)\ \ \ \ \ \ \ \ \ \ \ \ \mathbf{d}\Big(D^{(1)}_N\Big)=(1,0)\nn\\
\mathbf{d}\Big(\mathbf{W}_{n}^{(m)}\Big)&=&(mn,-n)
\ee
This puts the operators on the integer two-dimensional lattice, and the commutative families $\mathbf{W}_{-n}^{(m)}$ are just rays $(mn,-n)$ on this lattice, which makes the picture similar to that in the limiting case of the $W_{1+\infty}$ algebra. In fact, they are just halves of the corresponding Heisenberg subalgebras \cite{MMMP}. These Heisenberg subalgebras are associated with lines of an arbitrary rational slope $(rn,sn)$ for coprime $r$ and $s$ (also similarly to the $W_{1+\infty}$ algebra case), their manifest construction will be discussed elsewhere. Note that our choice of rays corresponds not to the rays in the DIM root lattice $e_{(n,m)}$, but to their linear combination. Thus, in terms of \cite{MMMP}, they are rather cones.

Actually, existence of the Heisenberg subalgebra associated with any rational slope follows from existence of the Miki $SL(2,\mathbb{Z})$-automorphism of the DIM algebra \cite{Miki1,Miki}, and the operator $\hat{\mathbf{O}}$ that we have constructed is nothing but the operator which generates this automorphism in the case of integer slopes.

\subsection{Operator $\hat{\mathbf{O}}$ in terms of Macdonald difference operators}
The operator $\hat{\mathbf{O}}$ can be also constructed directly from the Macdonald difference operators \cite[p.315, Eq.(3.4)]{Macdonald}
\begin{eqnarray}\label{Hamk}
	D_{N}^{(k)}&=&\sum_{1\leq i_1<\ldots<i_k \leq N} \left(\Delta_N^{-1}\prod_{m=1}^{k} T_{t,i_m}\cdot\Delta_N\right)\prod_{m=1}^{k} T_{q,i_m} \nonumber \\
	&=&\sum_{I\subseteq \left\{ 1,\cdots,N\right\}, |I|=k}t^{\binom{k}{2}}\prod_{i\in I,j\notin I} \frac{tx_i-x_j}{x_i-x_j}\prod_{i\in I}T_{q,i}.
\end{eqnarray}
The Macdonald polynomials $P_{\lambda}$ are eigenfunctions of (\ref{Hamk}) with some eigenvalues $\Lambda_\lambda^{(k)}$, \cite{Macdonald}
\begin{equation}
	D_{N}^{(k)} P_{\lambda}(x;q,t)=\Lambda_\lambda^{(k)} P_{\lambda}(x;q,t)
\end{equation}

Consider the generating function $D_{N}(Y)=\sum_{k=1}^N D_{N}^{(k)}Y^k$, then its action on the Macdonald polynomial is given by
\begin{eqnarray}
	D_{N}(Y) P_{\lambda}(x;q,t)
	=\prod_{i=1}^{N}\left(1+Yq^{\lambda_i}t^{N-i}\right)P_{\lambda}(x;q,t).
\end{eqnarray}

Now suppose that $|q|>1$ and consider the product
\begin{equation}
	\hat{\mathbf{O}}_{aux}(z):=\prod_{k=0}^\infty D_{N}(Y=-q^{-k}t^{-N}z)
\end{equation}
Using the identity
\begin{equation}
	\prod_{i=1}^N{1-zq^{\lambda_i}t^{-i}\over 1-zt^{-i}}=\prod_{(i,j)\in \lambda}{1-zq^j t^{-i}\over 1-zq^{j-1}t^{-i}}
\end{equation}
one can note that the eigenvalues of this operator are
\begin{equation}
	\prod_{i=1}^N\prod_{j=0}^\infty(1-zq^{-j}t^{-i})\cdot\prod_{(i,j)\in \lambda}(1-zq^{j}t^{-i}).
\end{equation}
Hence, we can finally construct the operator $\hat{\mathbf{O}}$:
\begin{equation}\label{opero}
\boxed{
	\hat{\mathbf{O}}:={(1-q)^{-\hat{E}}	\hat{\mathbf{O}}_{aux}(q^{-1}t^{N+1})\over \prod_{i=1}^N\prod_{j=1}^\infty(1-t^{N+1-i}q^{-j})}=
	(1-q)^{-\hat{E}}{\prod_{k=1}^\infty D_{N}(-q^{-k}t)\over \prod_{i=1}^N\prod_{j=1}^\infty(1-t^{N+1-i}q^{-j})}
}
\end{equation}
\\\\
In order to compare these formulas with those of the previous subsection, notice that, in Ref. \cite{Awata1}, H. Awata and H. Kanno expressed the Macdonald difference operators $D_N^{(k)}$ in terms of power sum variables $p_k=\sum_ix_i^k$. They defined the generating function
\begin{eqnarray}
	D_{N}(wt^{\frac{1}{2}-N})\exp\left\{\sum_{n>0}\frac{1}{n}\frac{(-wt^{{1 \over 2}-N})^n}{1-t^{n}} \right\}=:\sum_{r \geq 0}w^r H_N^{(r)}
\end{eqnarray}
i.e.
\begin{eqnarray}
	H_N^{(r)}=\sum_{s=0}^{\min(r,N)}t^{{s \over 2}-rN}e_{r-s}(t^{\rho})D_N^{(s)}, \quad r=0,1,\cdots,
\end{eqnarray}
$e_{k}(x)$ is the elementary symmetric function and $\rho=\left(-{1\over2},-{3\over2},-{5\over2},\cdots\right)$.

Under the limit $H^{(r)}=\lim_{N \to \infty}H_N^{(r)}$, one obtains the Hamiltonians acting on the $p_k$-variables:
\begin{eqnarray}
	H^{(r)}&=&e_r(t^{\rho})\oint\prod_{k=1}^{r}\frac{\mathrm{d}z_k}{2\pi iz_k}
	\Delta(z;t^{-1})\exp\left\{\sum_{n>0}\sum_{k=1}^{r}{1-t^{-n}\over n}z_k^n p_n\right\}\times \nonumber \\
	&&\times\exp\left\{\sum_{n>0}\sum_{k=1}^{r}(q^{n}-1)z_{k}^{-n} \frac{\partial}{\partial p_n}\right\},
\end{eqnarray}
where $\Delta(z;t)=\exp\left\{-\sum_{n>0}\sum_{1\leq i<j\leq r } {1-t^{n}\over n}(z_i/z_j)^n\right\} $.

\section{Discussion}

We have outlined the construction of the most basic type of $(q,t)$-deformed partition functions, which are a deformation of the skew Hurwitz $\tau$-functions.
We discussed the operators $\mathbf{W}_{-n}^{(m)}(q,t|\vec{N})$ which constitute a commutative family, and hence are Hamiltonians for some integrable system. The nature of this integrable system is yet to be understood.
We did not discuss an algebraic meaning of the constructed commutative  families and of the cut-and-join rotation operator $\hat{\mathbf{O}}(q,t|N)$, however, this subject deserves an attention. In particular, it has to include a $(q,t)$-deformation of commutation relations of the $W_{1+\infty}$-algebra. We have made only the first step: a realization in (\ref{qtW}) of the operator $\hat{\mathbf{O}}(q,t|N)$ and, hence, of $\mathbf{W}_{-n}^{(m)}(q,t|\vec{N})$ operators in terms of generators of the quantum toroidal algebra $U_{q,t^{-1}}(\hat{\hat{\mathfrak{gl}}}_1)$ in the Fock representation $\mathcal{F}_{q,t^{-1}}^{(1,0)}$.
Furthermore, we did not touch matrix model representations of
generic partition functions, which is also a subject for future work (see also \cite{MMZ}). It would be also interesting to clarify relations to $(q,t)$-deformations of integrable structures, like the ones considered in \cite{Awata2,Awata3,JE}.

\section *{Acknowledgments}

We are grateful to the referee of this paper for a series of valuable comments. This work is supported by the National Natural Science Foundation of China (Nos. 11875194 and 12205368),
by the grant of the Foundation for the Advancement of Theoretical Physics “BASIS” and by
the joint grant 21-51-46010-ST-a.


\end{document}